\documentclass[a4paper]{article}

\usepackage{INTERSPEECH2020}
\usepackage{multirow}
\usepackage{makecell}
\usepackage{color}
\usepackage{algorithm}
\usepackage{algorithmic}
\usepackage{graphicx}
\usepackage{lipsum}

\usepackage{xcolor}
\usepackage{tikz,subfigure}
\usetikzlibrary{automata, positioning, arrows}\usetikzlibrary{decorations.pathreplacing,positioning}
\definecolor{xablue}{HTML}{00BFFF}
\definecolor{xayellow}{HTML}{FACD00}
\definecolor{xapink}{HTML}{FFD700}
\definecolor{xagreen}{HTML}{7CFC00}
\definecolor{xagray}{HTML}{545454}
\definecolor{bb5e00}{HTML}{FFAF60}
\definecolor{ffe4ca}{HTML}{FFE4CA}
\definecolor{color66B3FF}{HTML}{CECEFF}
\pgfdeclarelayer{back}
\pgfdeclarelayer{front}
\pgfsetlayers{back,main,front}
\makeatletter
\pgfkeys{%
	/tikz/on layer/.code={
		\pgfonlayer{#1}\begingroup
		\aftergroup\endpgfonlayer
		\aftergroup\endgroup
	},
	/tikz/node on layer/.code={
		\gdef\node@@on@layer{%
			\setbox\tikz@tempbox=\hbox\bgroup\pgfonlayer{#1}\unhbox\tikz@tempbox\endpgfonlayer\egroup}
		\aftergroup\node@on@layer
	},
	/tikz/end node on layer/.code={
		\endpgfonlayer\endgroup\endgroup
	}
}
\def\node@on@layer{\aftergroup\node@@on@layer}
\makeatother
\newcommand{\kthree}{\tikz{\draw[step=0.1cm,very thin,fill=xablue] (0,0)  grid (0.1,0.3) rectangle (0,0);}}
\newcommand{\kfive}{\tikz{\draw[step=0.1cm,very thin,fill=xapink] (0,0)  grid (0.1,0.5) rectangle (0,0);}}
\newcommand{\kseven}{\tikz{\draw[step=0.1cm,very thin,fill=xagreen] (0,0)  grid (0.1,0.7) rectangle (0,0);}}
\newcommand{\kthreeh}{\tikz{\draw[step=0.1cm,very thin,fill=xablue] (0,0)  grid (0.3,0.1) rectangle (0,0);}}
\newcommand{\kfiveh}{\tikz{\draw[step=0.1cm,very thin,fill=xapink] (0,0)  grid (0.5,0.1) rectangle (0,0);}}
\newcommand{\ksevenh}{\tikz{\draw[step=0.1cm,very thin,fill=xagreen] (0,0)  grid (0.7,0.1) rectangle (0,0);}}

\newcounter{daggerfootnote}

\newcommand\blfootnote[1]{%
	\begingroup
	\renewcommand\thefootnote{}\footnote{#1}%
	\addtocounter{footnote}{-1}%
	\endgroup
}

\title{Neural Architecture Search on Acoustic Scene Classification}
\name{Jixiang Li$^{\dagger}$, Chuming Liang$^{\dagger}$, Bo Zhang, Zhao Wang, Fei Xiang, Xiangxiang Chu}
\address{Xiaomi AI Lab}
\email{\{lijixiang, liangchuming1, zhangbo11, wangzhao3, xiangfei, chuxiangxiang\}@xiaomi.com}

\begin{document}
	
	\maketitle
	%

	\begin{abstract}
		Convolutional neural networks are widely adopted in Acoustic Scene Classification (ASC) tasks, but they generally carry a heavy computational burden. In this work, we propose a high-performance yet lightweight baseline network inspired by MobileNetV2, which replaces square convolutional kernels with unidirectional ones to extract features alternately in temporal and frequency dimensions. Furthermore, we explore a dynamic architecture space built on the basis of the proposed baseline with the recent Neural Architecture Search (NAS) paradigm, which first train a supernet that incorporates all candidate architectures and then apply a well-known evolutionary algorithm NSGA-II to discover more efficient networks with higher accuracy and lower computational cost from the supernet. Experimental results demonstrate that our searched network is competent in ASC tasks, which achieves 90.3\% F1-score on the DCASE2018 task 5 evaluation set, marking a new state-of-the-art performance while saving 25\% of FLOPs compared to our baseline network.
	\end{abstract}
	\noindent\textbf{Index Terms}: acoustic scene classification, neural architecture search, deep learning, NSGA-II
	
	\blfootnote{Accepted to Interspeech 2020. $^\dagger$: Equal contribution. }
	
	\section{Introduction}
	Acoustic Scene Classification (ASC) is an important task  in the field of audio understanding and analysis, which classifies an audio stream into one of the predefined acoustic scenes. Recently, ASC has drawn increased attention from academy to  industry, due to its great potential in many applications like context aware devices \cite{eronen2005audio}, acoustic monitoring \cite{ntalampiras2009acoustic}, and assistive technologies \cite{bugalho2009detecting}. 
	In the early stages, many traditional machine learning methods such as GMM \cite{mesaros2016tut}, HMM \cite{eronen2005audio} and SVM \cite{mesaros2018detection} have been applied.
	Now with the great success of deep learning in computer vision and the availability of larger audio datasets, methods based on DNN \cite{mun2017deep}, CNN \cite{han2016acoustic} and RNN \cite{vu2016acoustic} have gradually been dominant in ASC. Among them, CNN-based methods have obtained the most state-of-the-art results because of their excellent capability of learning high-level features by exploiting time-frequency pattern of a signal \cite{mesaros2018acoustic}. For example, Inoue \cite{inoue2018domestic} ensembled four CNN models that championed in DCASE 2018 task 5 \cite{Dekkers2018_DCASE}. Hershey \cite{hershey2017cnn} compared different CNN architectures (VGG \cite{simonyan2014very}, Xception \cite{chollet2017xception} and ResNet \cite{he2016deep} etc.) in an audio classification task and all of them showed promising results. However, in vision tasks, most architectures above have been outperformed by advanced networks such as MobileNetV2 \cite{sandler2018mobilenetv2} in terms of the number of parameters and computational cost.
	
	To automatically get higher-performance networks than those designed by human experts, neural architecture search (NAS) has been adopted in many vision and NLP tasks, including our previous work on image super-resolution and image classification \cite{chu2019multi,chu2019fast,chu2019moga}. A few have applied NAS in speech like keyword spotting \cite{veniat2019stochastic,mazzawi2019improving}. We are thus motivated to validate NAS applicability in ASC task. Early NAS methods based on reinforcement learning \cite{zoph2018nasnet} and evolutionary algorithms \cite{real2019AmoebaNet} consume immense GPU resources to evaluate thousands of neural networks. Recent gradient-based methods  \cite{liu2019darts} require only a small amount of GPUs, but it is less robust and has weak reproducibility. On the contrary, the one-shot \cite{bender2018oneshot} paradigm is more promising mainly for three reasons: robustness to generate powerful architectures, moderate consumption of GPU resources, and convenience for solving multi-objective problems (MOPs, e.g. searching for a model with a trade-off between accuracy and computational cost). Generally, a supernet that comprises of all candidate architectures is constructed and trained fully  on a target dataset. Next, by inheriting weights directly from the supernet, candidates can have a good performance, from which we can judge competitive ones from the rest.   
	Beyond  \cite{bender2018oneshot}, single-path one-shot  \cite{guo2019single} and FairNAS \cite{chu2019fairnas} also exhibit improved training stability and more accurate ranking. 

	In this paper, we propose a lightweight network of high-performance inspired by MobileNetV2's inverted bottleneck block. On this basis, we also search for better architectures with the one-shot NAS doctrine. Many previous works in vision used CNNs as a feature extractor, mostly with square k$\times$k kernels since the information has the same nature in the spatial resolution. For ASC's mel-spectrogram, however, horizontal and vertical information has different implications, namely, temporal relation and frequency distribution. Square kernels are thus less appropriate. Based on the above consideration, we adapt MobileNetV2 with unidirectional (1$\times$k and k$\times$1) kernels to separately handle each dimension's information. Experiments in section \ref{section_experiment_baseline} demonstrate the proposed unidirectional convolution network has outstanding performance in DCASE 2018 Task 5  \cite{Dekkers2018_DCASE}. Further, we leverage fair supernet training strategy \cite{chu2019fairnas} and NSGA-II algorithm \cite{deb2002fast} to search for network architectures with higher accuracy and less computation. In section \ref{section_experiment_nas} we illustrate the details of the searched network that achieve state-of-the-art results on DCASE 2018 task 5.

	\section{Acoustic Scene Classification with Neural Architecture Search}
	\subsection{Proposed Baseline Network}
	
\begin{figure}[tb]
		\centering
		\begin{tikzpicture}[->,>=stealth',shorten >=1pt,auto,node distance=0.5cm,
		semithick,scale=0.7, every node/.style={scale=0.7}]
		\tikzstyle{cell} = [rectangle, minimum width=1cm, minimum height=0.2cm, font=\scriptsize,text centered, draw=black,rotate=90]
		\tikzstyle{block3} = [cell, fill=color66B3FF,draw=black]
		\tikzstyle{block5} = [cell, fill=color66B3FF,draw=black]
		\tikzstyle{GRU} = [cell, fill=bb5e00,draw=black]
		\tikzstyle{FC} = [cell, fill=ffe4ca,draw=black]
		\node[cell,label={[rotate=90,xshift=-0.6cm,yshift=0.2cm,font=\scriptsize]right:40*501*1}] (stem)  {Stem Conv K7$\times$7 S1$\times$1 F24};%
		\node[block3, below of=stem] (b1) {MB K3$\times$1 S2$\times$1 F32};
		\node[block3, below of=b1] (b2) {MB K3$\times$1 S1$\times$1 F32};
		\node[block3, below of=b1, xshift=0.1cm, yshift=-0.1cm, node on layer=back,label={[rotate=90,below,xshift=-1.6cm,font=\scriptsize]$\times$2}] (b3) {MB K3$\times$1 S1$\times$1 F32};
		\node[block3, below of=b2, yshift=-0.1cm] (b4) {MB K1$\times$3 S1$\times$2 F48};
		\node[block3, below of=b4] (b5) {MB K1$\times$3 S1$\times$1 F48};
		\node[block3, below of=b4, xshift=0.1cm, yshift=-0.1cm, node on layer=back,label={[rotate=90,below,xshift=-1.6cm,font=\scriptsize]$\times$2}] (b6) {MB K1$\times$3 S1$\times$1 F48};
		\node[block3, below of=b5, yshift=-0.1cm] (b7) {MB K3$\times$1 S2$\times$1 F64};
		\node[block3, below of=b7] (b8) {MB K3$\times$1 S1$\times$1 F64};
		\node[block3, below of=b7, xshift=0.1cm, yshift=-0.1cm, node on layer=back,label={[rotate=90,below,xshift=-1.6cm,font=\scriptsize]$\times$2}] (b9) {MB K3$\times$1 S1$\times$1 F64};
		\node[block3, below of=b8, yshift=-0.1cm] (b10) {MB K1$\times$3 S1$\times$2 F80};
		\node[block3, below of=b10] (b11) {MB K1$\times$3 S1$\times$1 F80};
		\node[block3, below of=b10, xshift=0.1cm, yshift=-0.1cm, node on layer=back,label={[rotate=90,below,xshift=-1.6cm,font=\scriptsize]$\times$2}] (b12) {MB K1$\times$3 S1$\times$1 F80};
		\node[block3, below of=b11, yshift=-0.1cm] (b13) {MB K3$\times$1 S2$\times$1 F96};
		\node[block3, below of=b13, node on layer=front] (b14) {MB K3$\times$1 S1$\times$1 F96};
		\node[block3, below of=b13, xshift=0.1cm, yshift=-0.1cm, node on layer=main] (b15) {MB K3$\times$1 S1$\times$1 F96};
		\node[block3, below of=b13, xshift=0.2cm, yshift=-0.2cm, node on layer=back,label={[rotate=90,below,xshift=-1.7cm,font=\scriptsize]$\times$3}] (b16) {MB K3$\times$1 S1$\times$1 F96};
		\node[block5, below of=b14, yshift=-0.2cm] (b17) {MB K1$\times$5 S1$\times$4 F112};
		\node[block3, below of=b17, node on layer=front] (b18) {MB K1$\times$3 S1$\times$1 F112};
		\node[block3, below of=b17, xshift=0.1cm, yshift=-0.1cm, node on layer=main] (b19) {MB K1$\times$3 S1$\times$1 F112};
		\node[block3, below of=b17, xshift=0.2cm, yshift=-0.2cm, node on layer=back,label={[rotate=90,below,xshift=-1.7cm,font=\scriptsize]$\times$3}] (b20) {MB K1$\times$3 S1$\times$1 F112};
		\node[cell, below of=b18, yshift=-0.2cm] (gconv) {Global Conv K5$\times$1 S1$\times$1 F128};
		\node[GRU, below of=gconv] (gru) {GRU (256)};
		\node[cell, below of=gru] (maxp) {MaxPooling K1$\times$4 S1$\times$4 (64)};
		\node[FC, below of=maxp] (fc1) {Flatten \& FC (512)};
		\node[FC, below of=fc1] (fc2) {FC (9)};
		\draw [-latex] (stem)+(-0.6cm,0) -- (stem);
		\end{tikzpicture}
		\caption{The proposed baseline adapted from MobileNetV2 bottleneck (MB) for ASC. Note K and S refer to kernel size and stride. F means filters. The expansion rate of MB is 6.} 
		\label{fig:feature-extractor-baseline}
		\vskip -0.2in
	\end{figure}
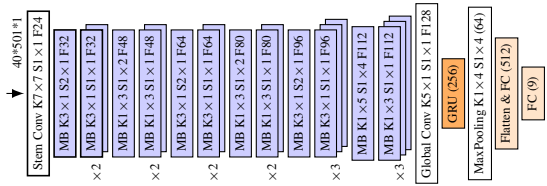

	In view of MobileNetV2's recent success in speech applications  \cite{luo2019conv,kriman2019quartznet}, we incorporate the MobileNetV2's bottleneck blocks  \cite{sandler2018mobilenetv2} as the core components in our baseline network. Our baseline mainly consists of three parts, namely a feature extractor (FE) module, a two-level gated recurrent unit (GRU) layer and two fully-connected (FC) layers. Specifically, the FE module comprises one stem convolution layer, 20 bottleneck blocks with 1$\times$1-depthwise-1$\times$1 convolution structure and one global frequency-dimensional convolution layer. We use unidirectional (1$\times$k and k$\times$1) kernels to replace original 3$\times$3 kernels in depthwise convolution of \cite{sandler2018mobilenetv2}. The configuration of this architecture is shown in Figure~\ref{fig:feature-extractor-baseline}. This FE module design is based on the following considerations. First, the residual property  \cite{he2016deep} of bottleneck block helps backpropagation and prevents vanishing gradients. Second, unidirectional kernels are utilized to extract features more attentively in temporal dimension and frequency dimensions. Besides, they reduce the number of parameters to prevent overfitting. Last, extracting features alternately in temporal and frequency dimensions is conducive to information flow and fusion.
	
	\subsection{Neural Architecture Search}
	
\begin{figure} [h]
		\vskip -0.1in
		\centering
		\includegraphics[scale=0.29]{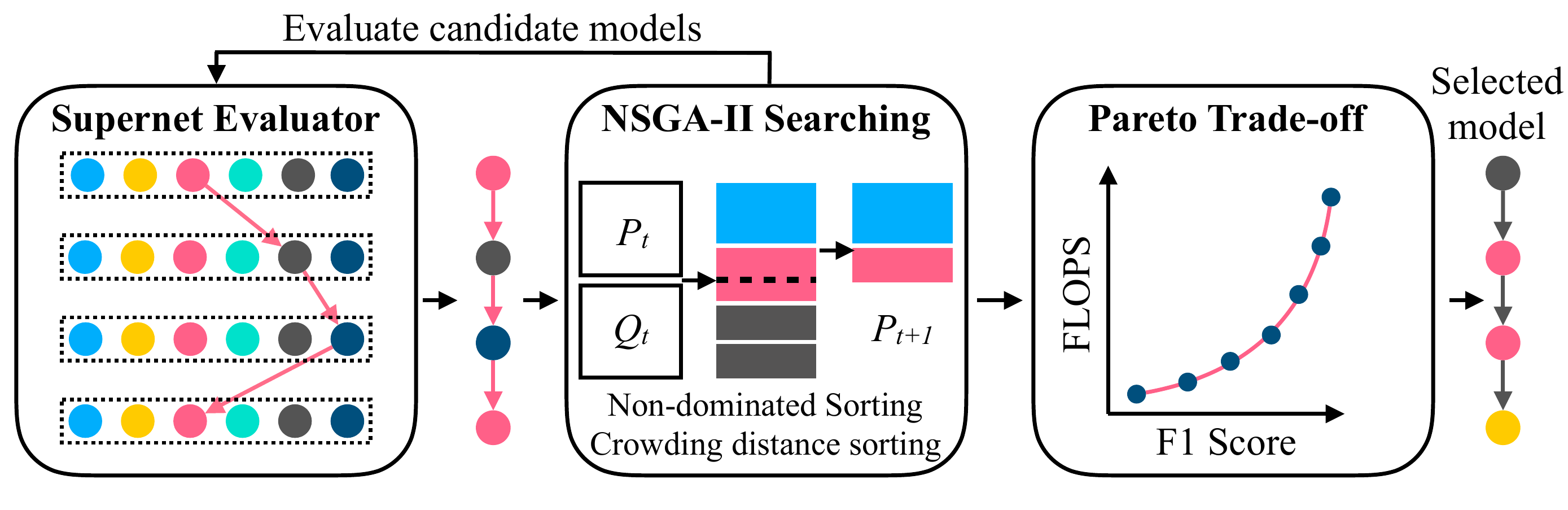}
		\vskip -0.1in
		\caption{The NAS pipeline consists of supernet training  \cite{chu2019fairnas} and NSGA-II Searching 	 \cite{deb2002fast}.} 
		\label{fig:nas-pipeline}
		\vskip -0.2in
	\end{figure}

	Based on our baseline network, we utilize NAS methodology to search for architectures with better performance. Concretely, we let our search targets be the kernel sizes and expansion rates in each inverted bottleneck block, which sums up to an enormous search space. Our supernet thus  consists of this searchable feature extractor along with the original GRU and FCs. We then train the supernet on the target dataset to obtain an evaluator for a good ranking among candidate models. Next, we use NSGA-II  \cite{deb2002fast}, an advanced evolutionary algorithm that features multi-objective optimization, to seek promising architectures with higher accuracy and less computation. Final competent architectures are selected from the Pareto-front obtained by NSGA-II and are trained from scratch. The overall NAS pipeline is illustrated in Figure~\ref{fig:nas-pipeline}.
	
	\subsubsection{Search Space}
	\label{section_ss}
	
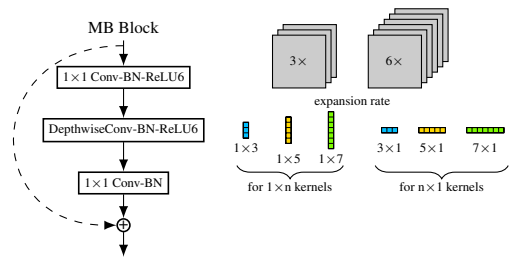
\begin{figure}[htb]
		\centering
		\begin{tikzpicture}[auto,node distance=0.8cm,
		semithick,scale=0.7, every node/.style={scale=0.7}]
		\tikzstyle{cell} = [rectangle, minimum width=1cm, minimum height=0.2cm, font=\scriptsize,text centered, draw=black]
		\tikzstyle{block} = [cell, fill=xablue,draw=black]
		\tikzstyle{feature} = [cell,minimum width=1cm, minimum height=1cm, fill=black!20!white]
		\node[cell,label={[above,yshift=0.5cm]MB Block}] (conv1)  {1$\times$1 Conv-BN-ReLU6};
		\node[cell, below of=conv1,yshift=-0.2cm] (dconv) {DepthwiseConv-BN-ReLU6};
		\node[cell, below of=dconv,yshift=-0.2cm] (conv2)  {1$\times$1 Conv-BN};
		\node[draw,circle, below of=conv2,inner sep=0pt] (skip) {+};
		\node [feature, right of=conv1,xshift=2.5cm, yshift=0.3cm,label={[above,xshift=1cm,yshift=-1.5cm, font=\scriptsize]expansion rate},node on layer=front] (ex3) {3$\times$};
		\node [feature, right of=conv1,xshift=2.6cm,yshift=0.4cm,node on layer=main] (ex3a) {3$\times$};
		\node [feature, right of=conv1,xshift=2.7cm,yshift=0.5cm,node on layer=back] (ex3b) {3$\times$};
		\node [feature, right of=ex3,xshift=1.5cm,yshift=0.5cm] (ex6) {6$\times$};
		\node [feature, right of=ex3,xshift=1.4cm,yshift=0.4cm] (ex6a) {6$\times$};
		\node [feature, right of=ex3,xshift=1.3cm,yshift=0.3cm] (ex6b) {6$\times$};
		\node [feature, right of=ex3,xshift=1.2cm,yshift=0.2cm] (ex6c) {6$\times$};
		\node [feature, right of=ex3,xshift=1.1cm,yshift=0.1cm] (ex6d) {6$\times$};
		\node [feature, right of=ex3,xshift=1cm] (ex6e) {6$\times$};
		\node [right of=dconv, xshift=1.5cm,label={[below,yshift=-0.4cm, font=\scriptsize]1$\times$3}] (k3) {\kthree};
		\node [right of=k3, label={[below,yshift=-0.7cm, font=\scriptsize]1$\times$5}]  (k5) {\kfive};
		\node [right of=k5, label={[below,yshift=-0.8cm, font=\scriptsize]1$\times$7}]  (k7) {\kseven};
		\node [right of=k7, xshift=0.3cm,label={[below,yshift=-0.3cm, font=\scriptsize]3$\times$1}] (k3h) {\kthreeh};
		\node [right of=k3h, label={[below,yshift=-0.3cm, font=\scriptsize]5$\times$1}]  (k5h) {\kfiveh};
		\node [right of=k5h, xshift=0.2cm, label={[below,yshift=-0.3cm, font=\scriptsize]7$\times$1}]  (k7h) {\ksevenh};
		\draw [latex-] (conv1) -- +(0,0.7cm);
		\draw [-latex] (conv1) -- (dconv);
		\draw [-latex] (dconv) -- (conv2);
		\draw [-latex] (conv2) -- (skip);
		\draw [-latex, dashed] (conv1)+(0,0.6cm) .. controls +(-180:2.7cm) and +(180:2.7cm) .. (skip);
		\draw [-latex] (skip) -- +(0,-0.6cm);
		\draw [decorate,decoration={brace,amplitude=5pt,mirror}] (k3.south west)+(0,-0.4cm) --+(2cm,-0.4cm) node [midway,below,yshift=-0.2cm, font=\scriptsize] {for 1$\times$n kernels};
		\draw [decorate,decoration={brace,amplitude=5pt,mirror}] (k3h.south west)+(0,-0.5cm) --+(2.5cm,-0.5cm) node [midway,below,yshift=-0.2cm, font=\scriptsize] {for n$\times$1 kernels};
		\end{tikzpicture}
		\caption{The searchable MobileNetV2's  bottleneck blocks. } 
		\label{fig:search-space}
		\vskip -0.1in
	\end{figure}

	We now describe the details of the search space for the FE module, where a certain number of bottleneck blocks with different settings can be chosen in each layer. We allow an expansion rate in \{3, 6\} for the first 1$\times$1 convolution and a kernel size in \{3, 5, 7\} for the depthwise convolution in each block as shown in Figure~\ref{fig:search-space}. For layer 17, we make an exception due to its large downsampling rate, where its expansion rate ranges in \{3, 6, 8\} and its kernel size in \{5, 7\}. The output filters of each block and the downsampling positions are fixed as in the baseline. In total, we have 20 searchable layers in the FE module, each with 6 choices of MB blocks. Hence, our search space contains $6^{20} \approx 10^{15}$ possible architectures. It is very hard to search for better architectures from such an enormous search space simply by trial-and-error. For convenience's sake, we refer the $\emph{c}$-th choice block $\emph{B}$ in layer $\emph{l}$ as $\emph{B}_{l}^{(\emph{c})}$, where $\emph{l} \in \{1,2, ..., 20\}$ and $\emph{c} \in \{1,2, ..., 6\}$.
	
	%
	\subsubsection{Supernet Training Strategy} 
	\label{section_supernet_training}
	
	We construct a supernet with the above searchable FE module, the original GRU and FCs together. To train the supernet, we use the same fairness strategy proposed in our previous work  \cite{chu2019fairnas}. Specifically, given a mini-batch of training data, instead of training the supernet as a whole, we \emph{uniformly sample without replacement} to have 6 models with no shared blocks for separate training, i.e. in the \emph{first} step we \textbf{randomly} select one choice from the 6 choices in each search layer to construct a model and the gradients of all parameters in this model are calculated by back-propagation. In the same way, in the \emph{second} step we \textbf{randomly} pick one choice from the \textbf{remaining unselected} choices in each layer to get another model and calculate its gradients. The rest may be deduced by analogy. After the \emph{sixth} step, \textbf{all} choices in each layer have been selected and their gradients are obtained. 
	Finally, we update the trainable parameters of the supernet according to the corresponding gradients altogether. The training algorithm is detailed in Algorithm \ref{algorithm_supernet_training}. Through this method, each choice has the same opportunity to update itself using every mini-batch of data, which stabilizes the training process.
	
	The trained supernet is then used to evaluate candidate models. In particular, we sample a candidate model from the supernet with its trained weights, and evaluate it on the validation set. In this way, we can quickly get the approximate performance of each candidate model in the supernet without extra training. 
	As proved in our previous work  \cite{chu2019fairnas}, the performance ranking among models obtained through this fair supernet training strategy is highly consistent with those trained from scratch.

	\begin{algorithm} [t]
		\caption{Supernet Training Strategy}
		\label{algorithm_supernet_training}
		\begin{algorithmic}
			\STATE {\bfseries Input:} training data loader $D_{train}$, the number $L$ of search layers, the number $C$ of choice blocks each layer, the number $I$ of training epochs, choice set $\Phi_l=\{1,2,3, ..., C\}$ for each search layer $l$ $\in \{1,2, ..., 20\}$
			\STATE {\bfseries Output:} the supernet with trained parameters
			\FOR{$i=1$ {\bfseries to} $I$}
			\FOR{data $d$, ground-truth $g$ {\bfseries in} $D_{train}$}
			\STATE clear gradients of all supernet parameters
			\FOR{$c=1$ {\bfseries to} $C$}
			\STATE initialize $\Psi_l=\Phi_l$ for each search layer if c == 1
			\FOR{$l=1$ {\bfseries to} $L$}
			\STATE randomly select an element \emph{e} from $\Psi_l$
			\STATE delete \emph{e} from $\Psi_l$
			\STATE get the choice $\emph{B}_{l}^{(\emph{e})}$ as $\emph{B}_{l}$ 
			\ENDFOR
			\STATE construct $\emph{model}_c$ by ($\emph{B}_{1}$, $\emph{B}_{2}$, ..., $\emph{B}_L$)
			\STATE calculate gradients $\nabla_c$ of $\emph{model}_c$ parameters by $d$, $g$
			\ENDFOR
			\STATE update all trainable parameters of the supernet by gradients $\nabla_1, \nabla_2, ..., \nabla_C$	
			\ENDFOR

			\ENDFOR
		\end{algorithmic}
	\end{algorithm}

	\subsubsection{Search Strategy}
	\label{section_search_stragegy}
	
	There are many ways to search, such as Random Search (RS), reinforcement learning (RL), and evolutionary algorithms (EA). Here we utilize an efficient evolutionary algorithm NSGA-II  \cite{deb2002fast} to search for promising architectures in the enormous search space. We adapt NSGA-II to our needs and only describe the differences in this section. Please refer to its original paper for the rest details (non-dominated sorting, crowding distance,  tournament selection, and Pareto-front, etc.). In this paper, an architecture is regarded as an \emph{individual} and encoded uniquely by one \emph{chromosome}. According to the illustration in Section \ref{section_ss}, we define the chromosome of an architecture as a list containing 20 elements corresponding to 20 searchable layers. Each  \emph{gene} in the chromosome is an integer ranges from 1 to 6 corresponding to 6 choices. Since we aim to search for architectures with higher accuracy and lower computation, we set the accuracy metric and computational cost as two \emph{objectives}. A \emph{population} of P=64 architectures are chosen for evolution which iterates for I=70 of generations. For crossover, we get two \emph{champions} (the architectures who defeat each opponent's architecture on \emph{objectives}) using tournament selection, then select two gene spots and swap genes at corresponding spots on these two champion chromosomes. As for mutation, we select one to four gene spots randomly and change gene values on these spots. 
	%
	In the early stage of the search, we set exploration ratio 100\% to explore enormous search space to find promising architectures by creating chromosomes randomly. As the search progressing, exploitation ratio $\alpha$, i.e. the number of individuals created by crossover and mutation / P, increases to 80\% gradually with the exploration ratio (1-$\alpha$) decreassing to 20\%, which pays more attention to searching for better ones near the searched promising architectures. The exploitation ratio $\alpha$ is defined as
	\begin{equation}
	\label{equation_exploitation_ratio}
	\left\{
	\begin{aligned}
	0 & , & i < 15, \\
	(i - 15) / 68.75 & , & 15 \leq i \leq 70
	\end{aligned}
	\right.
	\end{equation}
	where \emph{i} refers to the evolutionary iteration. The whole search algorithm is shown in Algorithm~ \ref{algorithm_search}.
	
	\begin{algorithm} [tb]
		\caption{Search Strategy}
		\label{algorithm_search}
		\begin{algorithmic}
			\STATE {\bfseries Input:} population size $P$, max iteration $I$, trained supernet weights $W$, mutation $\emph{m}$, crossover $\emph{c}$, tournament selection $\emph{ts}$ \cite{deb2002fast}, objective1 $\emph{obj1}$ for accuracy on validation set, objective2 $\emph{obj2}$ for FLOPs, non-dominated sorting and crowding distance sorting $\emph{nds}$-$\emph{cds}$ \cite{deb2002fast}
			\STATE {\bfseries Output:} the best population list
			\STATE initialize best population list $\emph{bp}$ = $\varnothing$
			\FOR{$i=1$ {\bfseries to} $I$}
			\STATE chromosome list $\emph{cl}$ = $\varnothing$, population list $\emph{p}$ = $\varnothing$
			\STATE update exploitation ratio $\alpha$ by Eq.~\ref{equation_exploitation_ratio}
			\STATE $\emph{cl}$ $\Leftarrow$ $\alpha$$\times$$P$ chromosomes by creating randomly
			\STATE $\emph{cl}$ $\Leftarrow$ (1-$\alpha$)$\times$$P$ chromosomes by $\emph{m}$($\emph{c}$($\emph{bp}$, \emph{ts}))
			\FOR{{\bfseries each} chromosome $\emph{c}$ {\bfseries in} $cl$}
			\STATE construct $\emph{model}$ by $c$ and $W$
			\STATE $\emph{accuracy}$ = $\emph{obj1}$($\emph{model}$)
			\STATE $\emph{flops}$ = $\emph{obj2}$($\emph{model}$)
			\STATE restore ($c$, $\emph{accuracy}$, $\emph{flops}$) into $\emph{p}$
			\ENDFOR 
			\STATE $\emph{p}$ $\Leftarrow$ $\emph{bp}$ $\cup$ $\emph{p}$
			\STATE $\emph{bp}$ $\Leftarrow$ Top$P$ Pareto-optimal individuals by $\emph{nds}$-$\emph{cds}$ on $\emph{p}$
			\ENDFOR
		\end{algorithmic}
	\end{algorithm}

	\section{Experiments and Analysis}
	
	\subsection{Dataset and Data Augmentation}
	\label{section_data_augment}
	
	We evaluate our models on the task 5 dataset of DCASE 2018 Challenge  \cite{Dekkers2018_DCASE} containing audios of 9 domestic activities. The whole dataset is divided into the development set (72984 segments) and the evaluation set (72972 segments). Each segment has four acoustic channels. We extract 40 log-mel band energies for each channel signal with a frame size of 40ms and hop size of 20ms, that gives us 40$\times$501 data matrix for each sample. 
	
	For data augmentation, we adopt the shuffling and mixing as in  \cite{inoue2018domestic} offline because of the highly imbalanced class distribution of the dataset. In particular, we increase the number of segments of minority classes (cooking, dishwashing, eating, other, social activity and vacuum cleaning) to 14K and the rest to 28K. Besides, Cutout  \cite{devries2017cutout} is used online when training to improve the regularization of models.
	
	\subsection{Training and Evaluation Setup}
	We train our models using Adam optimizer  \cite{kingma2014adam} with a weight decay of 1e-6 on 2 GPUs. We warm up the learning rate from 0 to 0.003 in the first three epochs and keep this learning rate in subsequent epochs. An exponential moving average with decay 0.9986 is used. The batch size is 192. 
	
	We evaluate our models on the entire evaluation set in two ways: \textbf{single}-mode and \textbf{ensemble}-mode. For single-mode, we just train one single model on augmented development set for 10 epochs, then evaluate it using the last checkpoint weights. For ensemble-mode, we use the official cross-validation folds  \cite{Dekkers2018_DCASE} and train four models on these four folds, respectively. As for scoring, we calculate macro-F1 score  \cite{Dekkers2018_DCASE} of each channel signal and then average them for the final score.  
	
	\subsection{Experiments on Baseline Network}
	\label{section_experiment_baseline}
	
	As an outperforming feature extractor, VGG  \cite{simonyan2014very} architecture is widely used by previous works like  \cite{Han2017vgglike} \cite{tanabe2018multichannel} \cite{Weimin2019attention} that achieve good performance, but its size and depth also bring the problem of high computational cost. As a comparison, we replace the FE module with original VGG16 while the experimental settings remain the same as our baseline to explore the relationship between the model capacity and the ability to extract features. In this section, we train our models in single-mode.
	
	\begin{table} [h]
		\caption{Performance comparisons with different feature extractor and input size.}
		\begin{center}
			\begin{footnotesize}
				\begin{tabular}{|*{2}{l|}*{3}{r|}}
					\hline
					\textbf{Model} & \textbf{Input} & \textbf{Params} & \textbf{FLOPs} & \textbf{F1 Score} \\
					& (H, W) & (M) & (G) & (\%) \\
					\hline
					\cite{Weimin2019attention}-ensemble & (64, 1250) & 26.65 & 17.98 &  89.1 \\
					\hline
					BaselineVGG   & \textbf{(40, 501)} & 17.24 & 5.99 &  89.4 \\
					Baseline   & \textbf{(40, 501)} & \textbf{3.31} & \textbf{2.03} &  \textbf{89.8} \\
					\hline
				\end{tabular}
			\end{footnotesize}
		\end{center}	
		\label{tab:comparison_baseline_and_VGGmodels}
		\vskip -0.2in
	\end{table}
	
	We can see from Table \ref{tab:comparison_baseline_and_VGGmodels} that our baseline model is 0.4\% F1 higher than the comparison model which just replaces the FE with original VGG16, while reducing 4G FLOPs and 14M parameters. It suggests that capacity and depth of model have no absolute relationship with the ability to extract features and demonstrates our proposed architecture with unidirectional kernels alternately in temporal dimension and frequency dimensions is efficient for the ASC task. Moreover, our baseline, as a single model with smaller input size, also outperforms  \cite{Weimin2019attention} which based on VGG architecture and ensembled by four models. The experiments in this section show that our proposed FE architecture can be used as a light-weight and high-performance feature extractor.
	
	\subsection{Experiments on NAS}
	\label{section_experiment_nas}
	
	\subsubsection{Search Result}

	In this section, we divide each category of the development set into a training set and a validation set in a 70\%:30\% ratio and augment the training set based on the strategy in section \ref{section_data_augment}. We train our supernet for 10 epochs on the augmented training set using section \ref{section_supernet_training} training strategy, then evaluate more than 4.4K candidate models on the validation set by the section \ref{section_search_stragegy} method. As a comparison, we also evaluate the same amount of candidates by RS instead of NSGA-II. 
	\begin{figure}[ht]
		\subfigure[]{
			\includegraphics[width=0.22\textwidth,scale=0.2]{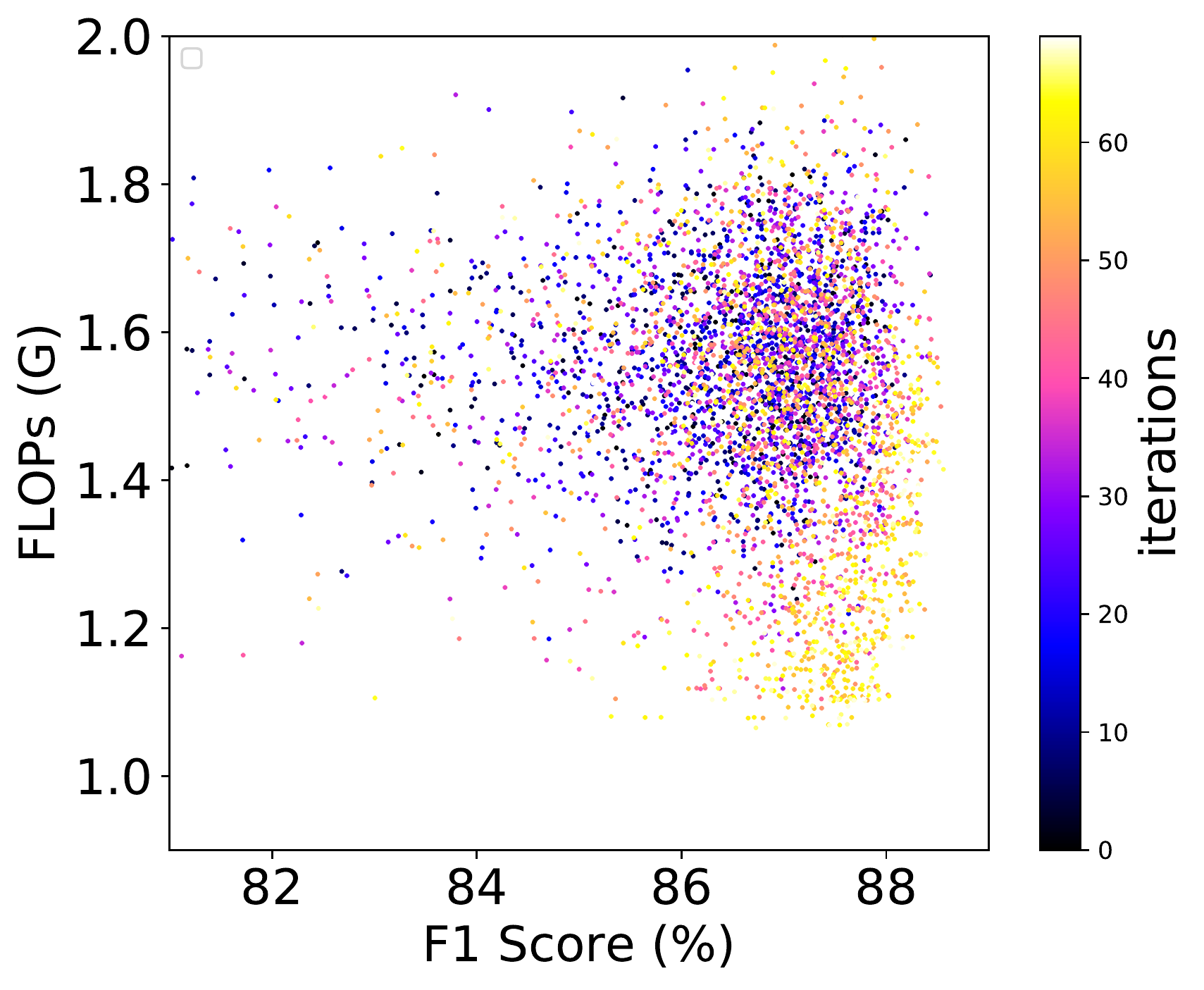}
		}
		\subfigure[]{
			\includegraphics[width=0.2\textwidth,scale=0.2]{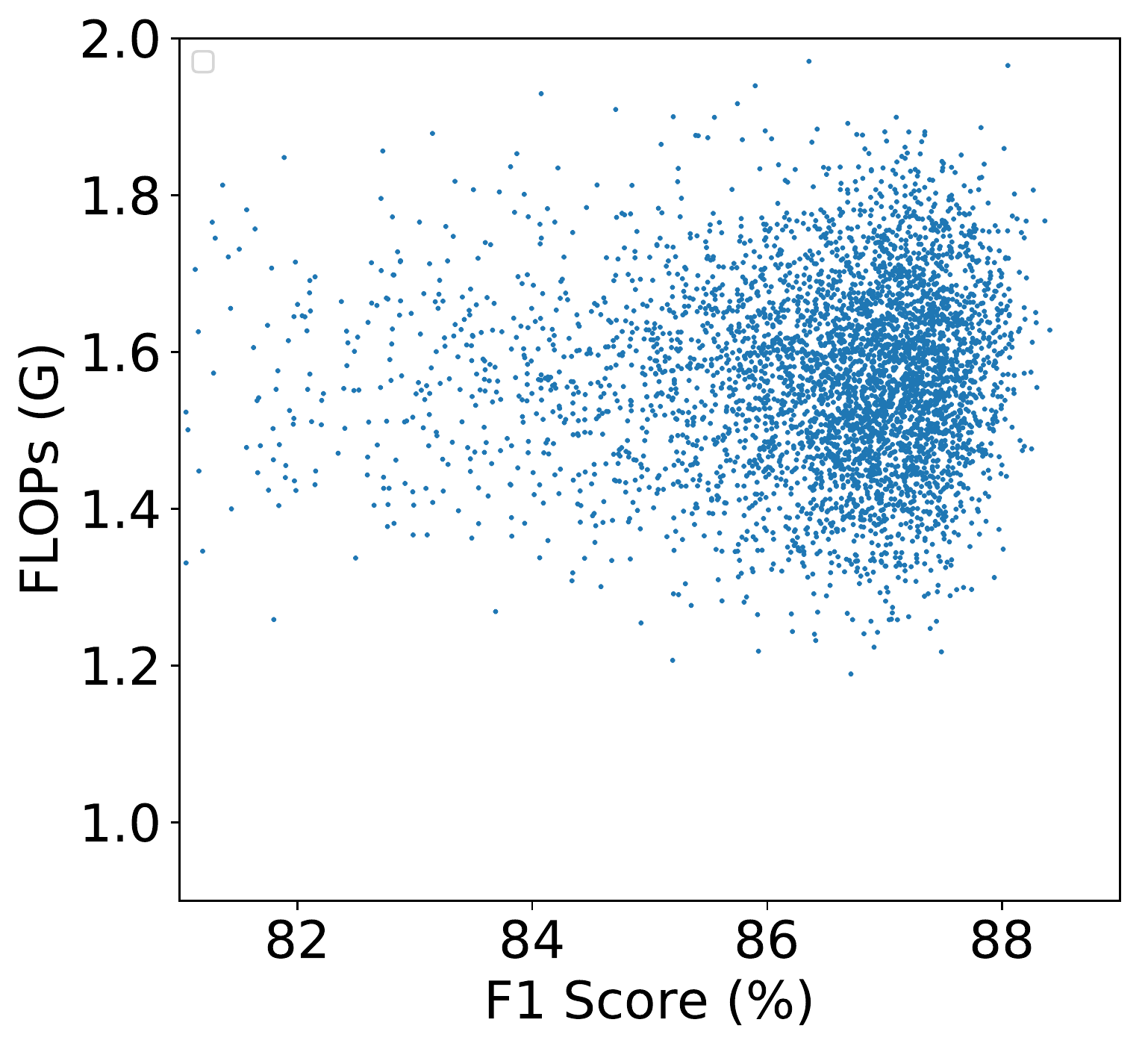}
		}
		\subfigure[]{
			\includegraphics[scale=0.28]{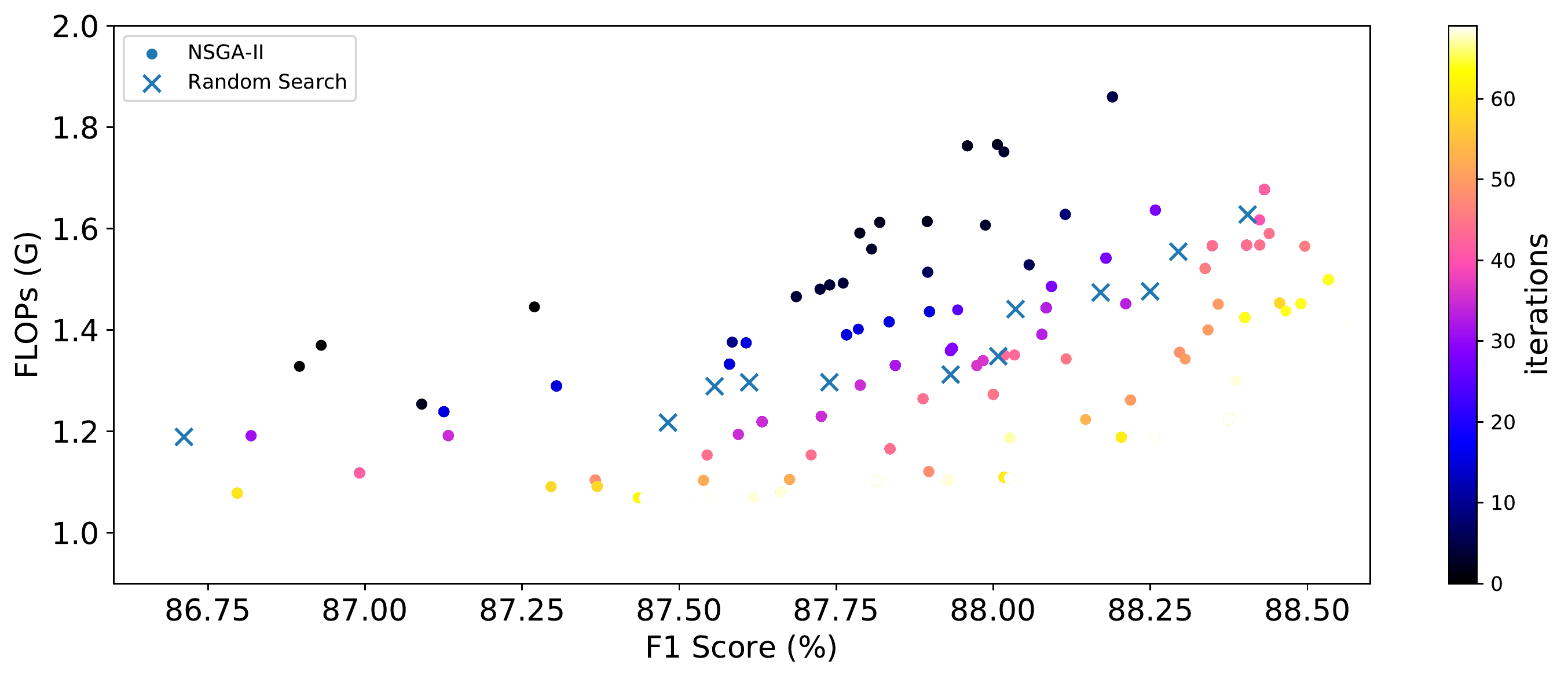}
		}
		\caption{(a) Models explored by NSGA-II (b) Random Search (c) Pareto-front of both methods.}
		\label{figure_searched_result}
	\end{figure}
	
	From Figure \ref{figure_searched_result} (a) we can see that in the early stage some promising models and lots of mediocre models are explored because of the large exploration rate. As the evolution iterates, the exploitation rate increases gradually, so that models near the searched promising models will be searched by crossover and mutation. In the later period of the evolution, due to the large exploitation ratio, more attention is invested in crossover and mutation so that more promising models are found. 
	The Figure~\ref{figure_searched_result} (b) clearly shows that the search distribution of RS only almost coincides with the \textbf{early} search area of NSGA-II, suggesting that RS is relatively weak on problems with multi-objective constraints.
	The Figure \ref{figure_searched_result} (c) shows the Pareto optimal models searched by NSGA-II and RS, which demonstrates that NSGA-II is more powerful.

	\subsubsection{Performance on DCASE 2018 Task 5}
	
	\begin{table} [ht]
		\caption{Class-wise performance comparison. Note \textbf{B-s} denotes the single baseline model, \textbf{N-s} denotes the single NASC-net model, \textbf{N-e} denotes the ensembled NASC-net model.}
		\begin{center}
			\begin{scriptsize}
				\begin{tabular}{|*{1}{l|}*{6}{r|}}
					\hline					\textbf{Class} &  \cite{Dekkers2018_DCASE}\footnotemark[1] &  \cite{inoue2018domestic}\footnotemark[1] &  \cite{Weimin2019attention} & \textbf{B-s} &  \textbf{N-s} &  \textbf{N-e} \\
					\hline
					Absence & 88.7 & \textbf{94.0} & 92.7 & 93.5 & 93.1 & 93.9 \\
					Cooking & 94.9 & 94.5 & 93.8 & 95.7 & \textbf{96.7} & 96.1  \\
					Dishwashing & 78.5 & 87.6 & 86.6 & 88.5 & \textbf{89.2} & 88.9  \\
					Eating & 81.7 & \textbf{88.7} & 88.0 & 88.0 & 88.5 & \textbf{88.7}  \\
					Other & 40.2 & 57.3 & 58.8 & 58.6 & 59.3 & \textbf{60.1}  \\
					Social activity & 96.5 & 97.2 & \textbf{97.9} & 97.4 & 97.0 & 97.8  \\
					Vacuum clean & 95.9 & 97.2 & 95.3 & 97.1 & 97.4 & \textbf{97.7}  \\
					Watching TV & 99.9 & \textbf{100.0} & \textbf{100.0} & \textbf{100.0} & \textbf{100.0} & \textbf{100.0}  \\
					Working & 81.5 & 89.4 & 88.4 & 89.3 & 88.9 & \textbf{89.7}  \\
					\hline
					F1 score & 84.2 & 89.5 & 89.1 & 89.8 & 90.0 & \textbf{90.3}  \\
					\hline
				\end{tabular}
			\end{scriptsize}
		\end{center}	
		\label{tab:comparison_all_models}
		\vskip -0.2in
	\end{table}
	
	\footnotetext[1]{Entire Evaluation Set of DCASE2018 task5 was divided into unknown microphone set (0.4286=3/7) and known microphone set (0.5714=4/7) for each category. Based on the ratio of the two sets, we can calculate the F1-score for entire Evaluation Set based on the official published leaderboard, e.g. the F1-score of category "Absence" in  \cite{Dekkers2018_DCASE} is 88.7\%=87.7\%$\times$0.4286+89.4\%$\times$0.5714.}
	
\begin{figure}[t]
		\centering
		\begin{tikzpicture}
		[->,>=stealth',shorten >=1pt,auto,node distance=0.5cm,semithick,scale=0.7, every node/.style={scale=0.7}]
		\tikzstyle{cell_e3} = [rectangle, text width=2.8cm, minimum height=0.2cm, font=\scriptsize,text centered, draw=black,rotate=90]
		\tikzstyle{cell_e6} = [rectangle, text width=3.2cm, minimum height=0.2cm, font=\scriptsize,text centered, draw=black,rotate=90]
		\tikzstyle{block3_e3} = [cell_e3, fill=xablue, draw=black]
		\tikzstyle{block5_e3} = [cell_e3, fill=xapink, draw=black]
		\tikzstyle{block7_e3} = [cell_e3, fill=xagreen, draw=black]	
		\tikzstyle{block3_e6} = [cell_e6, fill=xablue, draw=black]
		\tikzstyle{block5_e6} = [cell_e6, fill=xapink, draw=black]
		\tikzstyle{block7_e6} = [cell_e6, fill=xagreen, draw=black]	
		\node[block5_e6             ]  (b1)  {MB K5$\times$1 S2$\times$1 E6 F32};
		\node[block7_e6, below of=b1]  (b2)  {MB K7$\times$1 S1$\times$1 E6 F32};
		\node[block5_e3, below of=b2]  (b3)  {MB K5$\times$1 S1$\times$1 E3 F32};
		\node[block7_e6, below of=b3]  (b4)  {MB K1$\times$7 S1$\times$2 E6 F48};
		\node[block5_e6, below of=b4]  (b5)  {MB K1$\times$5 S1$\times$1 E6 F48};
		\node[block3_e3, below of=b5]  (b6)  {MB K1$\times$3 S1$\times$1 E3 F48};
		\node[block5_e3, below of=b6]  (b7)  {MB K5$\times$1 S2$\times$1 E3 F64};
		\node[block7_e3, below of=b7]  (b8)  {MB K7$\times$1 S1$\times$1 E3 F64};
		\node[block3_e3, below of=b8]  (b9)  {MB K3$\times$1 S1$\times$1 E3 F64};
		\node[block3_e3, below of=b9]  (b10) {MB K1$\times$3 S1$\times$2 E3 F80};
		\node[block3_e6, below of=b10] (b11) {MB K1$\times$3 S1$\times$1 E6 F80};
		\node[block5_e3, below of=b11] (b12) {MB K1$\times$5 S1$\times$1 E3 F80};
		\node[block5_e3, below of=b12] (b13) {MB K5$\times$1 S2$\times$1 E3 F96};
		\node[block5_e3, below of=b13] (b14) {MB K5$\times$1 S1$\times$1 E3 F96};
		\node[block3_e3, below of=b14] (b15) {MB K3$\times$1 S1$\times$1 E3 F96};
		\node[block3_e6, below of=b15] (b16) {MB K3$\times$1 S1$\times$1 E6 F96};
		\node[block5_e6, below of=b16] (b17) {MB K1$\times$5 S1$\times$4 E8 F112};
		\node[block3_e3, below of=b17] (b18) {MB K1$\times$3 S1$\times$1 E3 F112};
		\node[block7_e6, below of=b18] (b19) {MB K1$\times$7 S1$\times$1 E6 F112};
		\node[block5_e6, below of=b19] (b20) {MB K1$\times$5 S1$\times$1 E6 F112};		
		
		\end{tikzpicture}
		\caption{The searched part of the feature extractor module. Note E denotes the expansion rate.}
		\label{fig:searched-part}
	\end{figure}
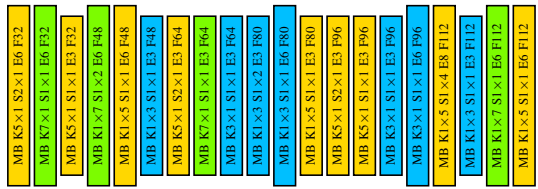

	Due to space limitation, we only select one searched model which has 1.53G FLOPs and 3.01M parameters (saving 25\% of FLOPs and 9\% of parameters compared to the baseline) on the Pareto-front of the 70-th iteration, though we can give lots of promising models after only one search process. We name this model \textbf{NASC-net} and the searched part of this model is illustrated in Figure~\ref{fig:searched-part}. NASC-net looks very weird in architectural perspective and it's almost impossible for human experts to design such an architecture, so we select it as a typical searched model and train it from scratch in single and ensemble mode, respectively. The performances on the evaluation set of DCASE 2018 task 5 is shown in Table \ref{tab:comparison_all_models}. NASC-net outperforms the baseline model in terms of macro F1 score, which is also more light-weight than those expert-designed. Besides, an ensemble of NASC-net uplifts F1-score further to 90.3\% at a 4$\times$ cost.

	\section{Conclusions}
	
	In this paper, we present a novel and efficient network for ASC tasks where its feature extractor is inspired by MobileNetV2. We show that our proposed network can achieve both high performance and low computation. Besides, on the basis of the proposed network, we apply neural architecture search to achieve a more sophisticated architecture using the fairness supernet training strategy and NSGA-II algorithm. Our searched network obtains a new state of the art on the DCASE2018 Task5 dataset with much lower computation. We can conclude that NAS is applicable in the field of ASC and potentially in other acoustics domains.
	
	
	\bibliographystyle{IEEEtran}
	
	\bibliography{mybib}
	
\end{document}